# Final models: A finalistic interpretation of statistical correlation

Dario Compagno

Department of Information and Communication

University Paris Nanterre

Nanterre 92000

France

Abstract

This paper aims to extend the framework of causal modelling to teleological explanations. It conceives final models as second-order models produced by interventions on first-order causal models. It shows why such formalisation permits us to realise a finalistic interpretation of statistical correlation which is compatible with its usual causal interpretation. Initially, the paper identifies some conceptual conditions for statistical teleological analysis, specifically involving interventions. Then, it describes an explanation procedure for action and presents one simple example of identifiable final models. Finally, it compares these results with what could be obtained within a purely causalist framework.

## 1. Two kinds of answers to the question *why?*

For Judea Pearl [1] science aims to answer the question *why?* This means going beyond observable data: explanation also demands *a priori* hypotheses, which can be formalised as assumptions of mathematical models. The rising importance of contemporary causal research is due to the fact that causal models can be *identified* from the data: there is a way to determine which causal representations of the world are better than others. This means that even if we cannot really 'see' causes, but only correlations among variables, we can still make causal inferences based on assumptions and empirical observations.

Are causal inferences the only ones that we can make starting from data and models? Actually, Pearl and other researchers in contemporary causal analysis only aim to answer a subset of *why* questions. These are all causal questions, for which the answer has the form of "what caused that effect?" Pearl offers a metaphor for causality by saying that some variables 'listen' to others, that is, that their values depend on those of other variables. And Pearl appears to think that the only correct interpretation which can be given to this 'listening' metaphor is a causal one. Valid answers to the question *why* point towards a relationship of 'listening' among variables, and this 'listening' is homogeneous, so all *why* questions are at the end of one only nature, which is interpreted by Pearl and many others as causal.

For Elisabeth Anscombe [2], instead, there are two distinct kinds of answers that can be given to the question *why*, and which cannot be reduced one to the other. One kind of answer is causal and the other teleological. Causal questions aim to find the causes of some effect, while teleological questions aim to find the ends of some action. For example, if I ask why the water in the pot is boiling, I could receive a causal answer asserting that it boils because the stove under it is on, or I can receive a teleological answer stating that it boils in order to cook noodles in it. The second answer does not imply that water is boiling spontaneously, and that this event is uncaused, but it does imply that some event can *also* be interpreted as an action, with some precise ends.

In the most popular contemporary philosophical view of action, derived from the work of Anscombe and Donald Davidson [3], the behaviour which is immediately realised by the agent functions as causal means for obtaining further effects, a subset of which should be recognized as being the ends of the action. An agent intends to bring about some effects by willingly realising some bodily movements which are causally connected to these effects ("in the good way", as Davidson precises, which means in ordinary situations which do not cover all conceivable exceptions). Therefore, to interpret an event as an action still implies that this event has causal effects, and that *part of these effects* constitute the proper answer to teleological questions.

This paper aims to begin modelling quantitatively such a perspective, allowing for the detection of intentions from observable data and assumptions. Particularly, we aim at extending the methodology developed by Pearl. This choice is based on the consideration that, as a matter of fact, Pearl's conceptual framework is already grounded on some notion of means and ends: more precisely, on an intervention notion of causation, which is in turn teleological. Starting from Pearl's work, there is only the need to differentiate intended and unintended effects to turn causal models into what we would like to call *final models*. This would allow for interpreting measurable correlations in an entirely new way, and final models could become a tool for answering questions in the human and social sciences which cannot be given an answer in terms of causes and effects.

Today statistics does not recognize the validity of any teleological interpretation of data. Every correlation among observed variables has to be explained causally. This idea could be linked to the classical principle of causation (Mill) or to what was more recently expressed by Hans Reichenbach [4]. For Reichenbach, whenever we observe some correlation between two variables *A* and *B*, either *A* causes *B*, or *B* causes *A*, or a third variable (confounder) causes both. This effectively appears to shield statistics from any non-causal form of explanation; at the same time, this assumption also permits causal inference. It is in fact thanks to this premise that the dependence relationships predicted by a causal model can be compared to those observed in the data (see below). However, if Reichenbach [4] is right, does this mean that only causal explanations of observed correlations are valid? In other terms, is there a way to formulate the concepts of action and intention so that teleological explanations can be valid, identifiable, and also compatible with causal ones? This paper aims to show that a framework for the empirical study of goal-oriented behaviour, intended as a means-ends relationship among variables and interventions on variables, can actually be developed starting from purely causal assumptions.

## 2. Interventions: The bridge between causes and ends

Contemporary causal analysis formalises causality as a mathematical relationship among variables. Causal models are the common tool for representing and studying causality. A simple causal model includes variables and binary relationships. A directed acyclic graph (DAG) can be used to graphically represent a causal model: its nodes stand for variables and its arrows for oriented causal relationships. A causal model can in some cases be *identified* from the data, which means that it can be stated whether the model is compatible or not with the observed correlations. This can be done because models put some constraints on the dependency and independency relations among variables. Therefore, if a certain model is valid, we expect to observe (or not to observe) some correlations among the variables it represents. In this perspective, a causal model is an interpretation of statistical correlations that is intended as a causal explanation.

The most general conceptual presuppositions for causal research–traceable to Reichenbach's work–are that causes are really at work in our world (generating observable measurements), and that they are the only possible explanation of the observed correlations. If these conditions were not met, it may appear that causal models cannot be identified. In fact, if our model implied some dependency, and the dependency was not due to anything that can be modelled in the model itself (that is, to some causal relationship), then there would be no guarantee for the validity of that model. As an example, if in my model a hot stove causes the water in a pot right above it to boil, I can explain the observed correlation between the hot stove and the boiling water. If the water was in fact not boiling with the stove on, I would have to change my model with a new one, in which different causal relationships end up accounting for the observed correlation. But if some invisible sprite could prevent water from boiling in arbitrary moments, then nothing would help me to build a good model of the world: I would not know when my model is not working because of misspecification and when this is due instead to the magical power of sprites. And still, interestingly, Pearl shows that the only way to build a solid methodology for identifying causal models is with the help of a sort of sprite, fairly large in size, capable of performing "small miracles" in kitchens and elsewhere.

The matter with the discovery of causal relationships is that they do not manifest themselves univocally in measurable correlations. Observed correlation between two variables *A* and *B* may in fact be spurious. Correlation is called spurious if it can be explained by a causal model, in which *A* and *B* are not connected by any directed path–that is, a path which can be walked by always following the direction of the arrows. I may explain the correlation between the sales of ice creams and the number of shark attacks without inserting an arc from the first to the other variable in my model, because the correlation between the two is actually due to a third variable (heat), which acts as a confounder. Correlation is instead not considered to be spurious if the model used to explain the data has a directed path between the two variables. Such as when I want to explain the boiling of water pots with reference to hot stoves. It is not correct, therefore, to simply talk about "causal correlations" versus "spurious correlations". This is because within a causalist framework, all explanations are necessarily causal. Even spurious correlation is causal at the end: it just requires taking into account the right variables and arcs in our model, connecting *A* and *B* in tortuous ways.

How can spurious correlations be discovered, from a methodological perspective? Contemporary causal analysis is based on a manipulation conception of causation. Such an approach attributes a central role to *interventions*, because it is only via interventions that one can test whether measurable correlation and hypothetical models meet or not. That of intervention is an extremely peculiar concept, which lies right on the border of causal explanation itself. In simple terms, the way to understand whether two generic variables *A* and *B* are or aren't causally related in a non-spurious way is to directly manipulate the value of *A* while at the same time observing what happens to *B*. Intervening on *A* means that its value is not due to anything else than the power of the agent who performs the intervention. So if we intervene on *A* we are effectively isolating it from any possible confounder influencing its value and also other values in our model. If the manipulation of *A* does not depend on any other variable in the causal model, then any cooccurrent observable change in *B* can be interpreted as causally depending on *A*.

There is a sort of trick at work here, which has to be understood in detail. In a few words: the relationship between *A* and *B* and the manipulation of *A* cannot be represented in the same model and do not even happen in the same world of reference. In one world of reference, which is the 'natural' world we aim to explain with our causal analysis, there was no intervention and *A* was forced to follow all and only causal laws, its value be assigned by all confounders present in the model; while in the 'manipulated' world, *A* and only *A* has been intervened on, and so was free to take whatever value, while all the other variables remain causally bounded. The shift between these two worlds of reference is what permits causal analysis ("why did *B* change, because of which cause?") and it is also what we aim to explain with teleological analysis ("why was *A* manipulated, because of which intended effect?"). A scientist turns a stove on and off at random, but *why* does she do that? The concept of intervention strictly prevents any answer to this question–at least from within a purely causal perspective.

In simpler terms, Pearl explains the difference between the natural (causal) attribution of values to variables and the artificial (interventionist) attribution by contrasting the verbs *to see* and *to do*. To observe our arm rising is not the same thing as raising our arm, as Anscombe wrote [2]. In the first case, I am the spectator of causal chains, while in the second, I observe the world from a different perspective, that of an agent intervening in it. The English verbs *rise* and *raise* in the former example show well the distinction between the fact of observing a causal system at work, and that of "breaking" such a causal system, by freely assigning a value to some variables independently of their causes in the system *and* also at the same time observing the normal doing of subsequent causal relationships. This "small miracle" is a requirement of causal research, and it is also at the same time the way to conceive action we endorse in this paper.

To *see A* and to *do A* are not the same thing and are not given the same formal representation by Pearl, who developed a useful formalisation of the concept of intervention. Some variable *A* on which one intervenes is named *do(A)* in probabilistic formulas. So, the conditional probability under simple observation *P(B)|A* is not the same thing as the conditional probability under intervention *P(B)|do(A)*. Practically, in the first case we would be filtering some observed values of *A*, while in the second case all the occurrences of the variable would be assigned some arbitrary value. With an example, if I only looked at those days in which many ice creams are sold, I could also observe many shark attacks, because the days I retained are most certainly summer days. In the actual world these two events correlate, but they are not one the cause of the other. However, if I artificially started to give away ice creams all along the year, I would not subsequently observe an increased number of shark attacks, given that sharks would still find little swimmers when it's cold. The world of reference including such an intervention in it would then not be the real world, as no-one but me gives away ice creams all along the year. Such a world would very well be real and purely causal if it wasn't for me, acting in it and so breaking an absolutely real and perfectly observable correlation between ice cream sales and shark attacks.

A little paradoxically, it seems therefore that without human intervention nothing can be said of a world in which there is no human intervention[1]. The important word in the last sentence is not 'human', of course, but 'intervention'. Within the contemporary framework in causal research, interventions are welcome as tools for modelling but not as objects of modelling. Now, we argue that we can also say something about the worlds in which one assumes the existence of (human) intervention: an occasional breaking of causal laws, producing worlds which are not totally real or "natural", but that can still be inquired into. Particularly, we argue that action can be explained by measuring and studying the real effects produced by interventions. Teleological explanation puts the causes of action into parenthesis and inquiries into its effects.

---

[1] This idea was expressed by von Wright [5].

In causal analysis, the most notable consequence of intervening on a variable *A* is that this variable is freed ("becomes uncaused") by any other variable in the model. This is what permits us to unmask confounders and interpret correlation as spurious or not. Graphically, we remove from our model all inbound arrows into variable *A*. If variable *A* was endogenous in the original model (if it 'listened' to other variables in the model), now in the new "surgically modified" model it becomes exogenous: its value is determined by something which is not captured anymore by the model. This is the core strategy of contemporary causal research. Causal analysis means to study some variable which has some causes in the actual world and in the model of interest via a tool (intervention) that permits to mathematically cancel out its causes and to only care about its effects. Strategy which works also as a modelling of action.

Causal analysis, notably as developed by Pearl, requires this notion of "freeing a variable" (or "becoming uncaused"), both as a conceptual requirement and as its main empirical research strategy. "Becoming uncaused" is not the same thing as "being uncaused". The causal research framework is authentically causalist and respects the principle enounced by Reichenbach: no variable is uncaused in the world represented by a model without interventions. But the discovery of causes demands a temporary anti-causal assumption and a step into worlds in which people give away ice creams all along the year *in order to* study the effects of their real sales, particularly on sharks' behaviour. It is this assumption that allows to model goal-oriented behaviour empirically, that is, to give a teleological interpretation to statistical correlation, and so to identify "in order to" relationships.

### 3. M*-models (not causal anymore but not final yet)

Before introducing what we call "final models", let us evaluate what has already been done in causal research without receiving attention from the correct angle. Following Pearl, interventions produce modifications of the causal models which a researcher aims to identify. If we call *M* the model according to which we believe our observations work, by intervening on any variable in *M* we produce a derived model *M\**, in which one variable has been intervened on. For clarity, in this paper we generalise and call any causal model an *M-model* and any model obtained via intervention an *M\*-model*. We do not simply call "final models" these M*-models, because to obtain final models one more step has to be taken, one operation on M*-models which we will present below in §4. There is generally a many-to-many relationship between M-models and M*-models, as one can intervene on a number of variables, therefore producing different M*-models, and as one and the same M*-model could have been produced by several M-models and interventions. But given an M-model and one intervention, we obtain a unique M*-model as a result.

We know what M-models are, but what are M*-models? They are not M-models. Specifically, they are not smaller M-models. An M-model is an *a priori* assumption about the causal mechanism generating the observed data. Any conceivable variable in the world, if we follow Reichenbach, can be modelled by an M-model. Some variables can be exogenous in a valid M-model, meaning that their causes are not taken into account by the model itself. Still this does not mean that the exogenous variables are really, intrinsically or essentially exogenous (that they "are uncaused"). In fact, a larger and equally valid M-model could include some more variables, and as a result the exogenous variables of our original M-model would now show their causes in a larger M-model (their nodes in the larger graph would now have parents). The opposite way around, if I produce a smaller and still valid M-model removing variables from a larger M-model, now some variables may not have any parent, while they would still be supposed to keep respecting the principle of general causation. Smaller and larger valid M-models refer to the same world, and the exogenous nature of some variables has to be interpreted only locally (as a limitation of causal knowledge, not of fact).

Producing a smaller M-model by removing some variables does not produce an M*-model: an M*-model is not the pruning of an M-model. This because, firstly, in the M*-model only some arrows are removed (those pointing to the manipulated variable) and never any node. Generally, if I obtain an M*-model from a valid M-model, then the resulting M*-model is not a smaller and still valid M-model. It is easy to see why: by definition I removed some arrows that were surely kind of important for causally explaining reality. If I start selling ice creams in the winter, then I am generating data that should not be there to begin with, and the M*-model derived by my intervention is not a valid causal model. Summer is a thing, while the M*-model produced by intervention is not anymore causally valid. The confounders blocked by intervention are meaningfully absent only in an M*-model, and should instead be present in any valid and large enough M-model including them and their effects. If we remove the effect of a confounder on the cause we are studying, we are temporarily abandoning the real, empirical word. No M*-model can therefore ever truthfully represent reality, in a purely causalist perspective.

Any M*-model is the result of a "small miracle" (Lewis) [6], or more simply the conceivable result of an intentional action realised by an agent. This sentence cannot be translated–that is, *reduced*–into any sentence which makes sense in a purely causalist conceptual framework. In simpler words, actions are not facts or events, and cannot be described by first-order variables like *A* or *B* in a causal model. If it was possible to do so, we should be able to create an M-model perfectly equivalent to our M*-model, without any concept of intervention involved. With an example, we should produce an M-model in which ice cream sales and shark attacks are at the same time dependent and independent from each other (the common cause of hot weather must and also must not be part of the explanation). Hot weather correlates with people selling more ice cream but not with *me* selling more ice cream. This is because, in a language without the distinction between *to see* and *to do*, actions are just events or facts, and everything follows causal connections, so no variable can "become uncaused", freed from its causal influences, not even temporarily. For a purely observational machine, there is nothing like a spurious correlation between ice cream sales and shark attacks, but just a correlation which is as real as any other. Causality makes sense only if we can go beyond mere correlation, and this is also what permits teleological interpretation.

Which kind of world are we representing with an M*-model, if we stated that it cannot be a world in which causal relationships are the only reason for the attribution of values to variables? A world represented by an M*-model is peculiar. It is an ephemeral world in which all causal laws existing in nature are valid, but there is also an extremely limited number of free actions: namely one, the intervention producing the M*-model. Actions happen one world at a time, so to speak. This is because in a causal model, as we stated above, all the indefinite number of variables not included in the model are still considered to potentially exist empirically and potentially play a causal role[2]. So everything but the intervention has to follow causation without exception. Therefore, two interventions on two different variables in the same M-model do not happen in the same world of reference, because each one is bounded in the world of reference in which the other is free.

Let us restate that M*-models are already an implicit but essential part of the contemporary framework in causal research. We are just giving them a name to put them to the foreground and asking explicitly what they are, and if we can use them for something more. In order to do this, we need to turn M*-models into models capables of explaining interventions, that is, final models.

### 4. Final Models

We have already introduced almost all the basic conceptual resources to formalise teleological questions and answers, within the language of causal research, which now needs to be only slightly extended. A teleological question asks about the ends of an action.

---

[2] The world of reference of causal models is 'thin', in the sense of Ryle [7].

"Why is the water boiling", asks to be answered teleologically with something like "In order to cook noodles in it". This kind of question can be formalised by asking which variables determine or probabilistically influence the value of *do(A)* in a *final model*, where *do(A)* is the action to be explained. Final models are casual models with two additions: interventions and intentional effects.

Firstly, the *interventions* have to be explicitly tagged with *do*, and this also in the DAG's representing the models. In causal analysis, interventions are operations on models and never appear as variables themselves (no node in a DAG is ever tagged with the operator *do*). In teleological analysis, these operations on models are instead the object of research, and they have to appear as nodes of final models. This is why teleological analysis can be seen as a second-order form of analysis, with respect to what we call causal analysis at first-order: we aim to make assertions about models, which in turn make assertions about observable variables[3].

Teleological questions cannot be asked within a causal model. Either some variable *A* has some cause, within the model, or it hasn't. But to answer a teleological question, we need to make assertions about interventions of the form *do(A)*. This doesn't mean that the causal relationships on which actions are grounded cease to work: they are just amended in one point, the one which we consider as the "unnatural" or "freed" variable we intervened on. In other terms, *A* is not equivalent to *do(A)*, as we already knew. If I ask "why are you boiling water?", and I expect a teleological answer, I have a legitimate but secondary interest in the fact that boiling water (*A*) causes the cooking of noodles (*B*). My primary interest is in why you intervened in the world, by intentionally affecting an ongoing causal system by the means of *do(A)*. The answer *B*, "I am going to cook noodles in it" requires, conceptually and mathematically, to make assertions on the *do*'s that are affecting the natural development of the world–while keeping into place our first-level causal knowledge, without which action cannot be understood. So from a statistical perspective we need to empirically validate assertions about interventions.

---

[3] Final modelling is of higher-order compared to causal modelling because its object is not a variable *A* within an M-model, but the couple of models *M* and *M\** related by the intervention *do(A)*. In fact, *do(A)* is not simply an operation on the variable *A*, as it may appear from its written representation, but an operation on the model *M* in which *A* appears; that is, *do(A)* means different things in different *M-models*, because it removes different arcs. The conditional probability *P(B)|do(A)* should be understood as *P(B) in M\**, because what is really "made" by the intervention is a new, temporary world in which causal relationships are amended in one point.

Second and last thing needed in order to turn a causal model into a final model, beyond tagging one variable as intervention, is to also identify which variables we believe to be the *intentional effects* of the intervention. Practically, this means that in the final model we will *reverse the direction of the arrows* linking the intervened variable *do(A)* to some of its effects. The sentence "*B* was the aim of the action *A*" is represented in a final DAG by *do(A)* and by an arc going from B to *do(A)*. This is because *do(A)* 'listens' to *some* of its effects, and the agent performs *do(A)* in order to obtain *B*. Example: let our first-order causal model be *A causes B*, in which *A* stands for boiling water (the *fact* that some water is boiling) and *B* for cooked noodles. Now, the sentence "She boiled water in order to cook noodles", can be represented by a final model in which do(*A*), boiling water (the *act* of making some water boil), has an arc coming from *B*, cooked noodles. In this context, the aim of statistical teleological analysis is to identify this final model, that is, to state whether *B* was or not an intended effect of *do(A)*. Is it true or not that she boiled water in order to cook noodles? More generally, is it true or not that people in our population of interest boil water in order to cook noodles?

Final models are based on one general assumption: every action has some intended effect, and not all the effects of any action are intended. Teleological analysis as we conceive it in this paper is basically a way to differentiate intended and unintended effects starting from observational (or experimental, if available) data. There is a second conceptual requirement of teleological analysis: the intervened variable has to be under the total control of the agent. That is, *do(A)* does not have any parent node causally determining its value, before reversing the direction of the arcs linking it to its effects. This is already a standard requirement of causal analysis (producing what we called M*-models), so we will not spend more time on it.

## 5. One toy example of identifiable final models

Let us take into consideration one toy example of an identifiable final model. Our model describes some aspects of a room and includes four variables: *H* represents the state of an heating system (0 for off and 1 for on); *T* represents room temperature (0 for cold, 1 for warm, and 2 for hot); *W* represents the weather conditions (0 for bad and 1 for good); *B* represents the amount of the electric bill (0 for inexpensive and 1 for expensive). Firstly, let us represent a causal model $M_1$ depicting the causal relationships among these variables in Figure 1.

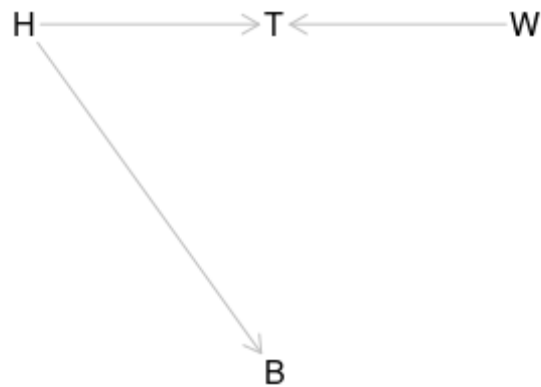

Figure 1: Graph representing the causal model $M_1$

In the world described by $M_1$ the state of the heating system impacts both the temperature in the room and the electrical bill. The room temperature is also influenced by the external weather. There is no other direct causal relationship. Some easy structural equations may describe this minimalistic world more in detail; for simplicity, let's assume that each exogenous variable (without parent nodes) may take the value of 0 or 1, and that each endogenous variable (with parent nodes) takes as value the sum of its parents' values.
The potential states of the world in which the causal constraints of $M_1$ hold are just four:
- either the weather is bad, the heating is off, the room is cold and the bill is cheap;
- or the weather is good, the heating is off, the room is warm and the bill is cheap;

- or the weather is bad, the heating is on, the room is warm and the bill is expensive;
- or the weather is good, the heating is on, the room is hot and the bill is expensive.

Let's then represent all the possible combinations of our variables with a table (Table 1).

| W | H | T | B |
|---|---|---|---|
| 0 | 0 | 0 | 0 |
| 1 | 0 | 1 | 0 |
| 0 | 1 | 1 | 1 |
| 1 | 1 | 2 | 1 |

Table 1: Possible combinations of values in the world modelled by $M_1$

In this extensional perspective, causal relationships are shown by the fact that some lines are *missing* from the table. Were we to observe any different combination, then we would have to revise our model: for example, if the temperature in the room was cold despite the heating being turned on. Causality is therefore a constraint on the potential states of this very small world. Some conceivable states of the world are causally impossible, and that is why we never expect to observe them. Importantly, W and H are *independent* of each other: independently of the weather being good or bad, we can observe the heating either on or off, and vice versa.

Now let us produce a final model from $M_1$, in which we consider H as the variable upon which the agent intervenes. We replace the node H in M with the node *do(H)*, indicating that all parents of H should be removed (in this case H had no parents in M). Our hypothesis for this teleological model $M_1$* is that the agent aims to bring about some change in the room's temperature. Therefore we invert the direction of the arrow connecting T and *do(H)*. Figure 2 depicts this situation.

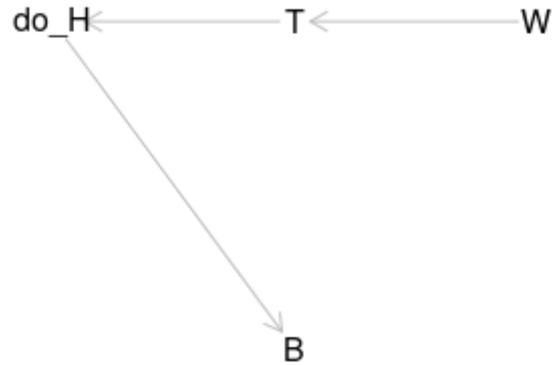

Figure 2: Graph representing the model $M_1^*$

Our aim is to test whether this model holds or not, through the analysis of observational (or experimental, if available) data. Discursively, we expect that the electrical bill is no issue for the agent, and that it plays no role in the decision of turning the heating on or not. We also expect that if the temperature is already warm, because of the external temperature, the agent will refrain from turning the heating on. This means that *we expect more potential states of the world to be missing*. If causality is a first constraint on the number of potential states of the world, finality is a further constraint adding up to the first. *To do* means to reduce the number of potential states that can become actual. If I do not want to feel cold, my action will prevent the cold potential states of the world from happening. That is, some worlds which would be causally possible, are not teleologically possible if the agent acts in view of impacting the room temperature, and so if *do(H)* 'listens' to the state of *T*. In the case of $M_1^*$ its conditions are identifiable: we can make a teleological inference and test whether or not this model holds. Figure 2 shows all and only the potential states of the world compatible with the hypothesis that the agent intervening on *H* aims to obtain a warm temperature in the room, that is, aims to causally set the value of *T* exactly to 1.

| W | H | T | B |
|---|---|---|---|
| 0 | 1 | 1 | 0 |
| 1 | 0 | 1 | 1 |

Table 2: The potential states of the world compatible with the goal *T = 1* in $M_1$*

Discursively, if we make the hypothesis that the agent intervenes in the causal system because he or she aims to obtain warm temperature, then we never expect to observe at the same time good weather and the heating on. This means that, if our final model holds, now the measurements of *W* and *H* must be *dependent* on each other, and this despite their first-level causal independence, that we noticed in $M_1$. The state of the heating now depends on ('listens' to) the weather: the agent turns the heating on if and only if the weather is bad. Action is a new constraint into the world and this constraint does not contradict the valid causal relationships connecting *H, W*, and *T*–it just adds further constraints based on *do(H)*. The dependance between *W* and *H* is measurable, and our teleological model is identifiable. Were we to observe the heating on during a sunny day, our teleological explanation would prove wrong. It should be noted that a graphical analysis of Figure 2 would allow us to predict the new dependence of *W* and *H* (because they are links of a so-called chain). Let us compare the different conditions we expect to observe under three different hypotheses about the goals of the agent: warm temperature (T = 1), never hot temperature (T < 2) and never cold temperature (T > 0). Table 3 compares the potential states of the world compatible with these hypotheses.

| W | H | T | B | *T = 1* | *T < 2* | *T > 0* |
|---|---|---|---|---|---|---|
| 0 | 0 | 0 | 0 | | V | |
| 1 | 0 | 1 | 0 | V | V | V |
| 0 | 1 | 1 | 1 | V | V | V |
| 1 | 1 | 2 | 1 | | | V |

Table 3: The potential states of the world according to three different goals in $M_1^*$.

In the case of $M_1^*$ the three goals all have different observation conditions, so they can be distinguished just with observational data. Globally, all three have the same presupposition: that *W* and *H* are dependent on each other. So to observe a dependency between these two variables would anyway suffice to identify the model $M_1^*$.

What if instead the agent intervened on the state of the heating for another reason? This should be explained with another, different final model. In our toy exemple, there is only another possible teleological explanation accountable by any model *M\** produced by *do(H)* in $M_1$, namely that the agent aimed to have an impact on the electrical bill. The intervention *do(H)* would therefore 'listen' to *B* and not anymore to *T*. Figure 3 depicts this alternative model $M_2^*$.

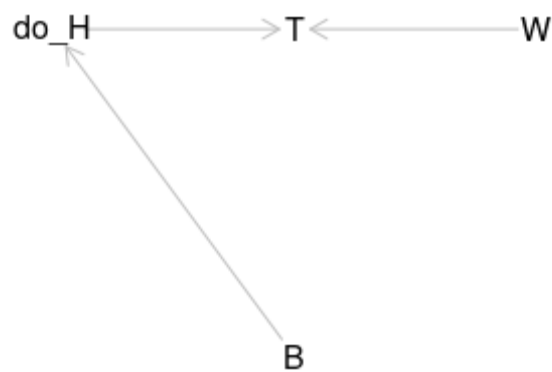

Figure 3: Graph representing the model $M_2^*$

We do not expect to observe the same potential states of the world as before if this second teleological model holds. If the agent only cares about his or her electrical bill, it is exclusively this variable B that *do(H)* will 'listen' to. Table 4 shows the potential states of the world which are compatible with the two potential goals of the agent, namely that B = 0 or that B = 1.

| W | H | T | B | *B = 0* | *B = 1* |
|---|---|---|---|---|---|
| 0 | 0 | 0 | 0 | V | |
| 1 | 0 | 1 | 0 | V | |
| 0 | 1 | 1 | 1 | | V |
| 1 | 1 | 2 | 1 | | V |

Table 4: The potential states of the world according to two different goals in $M_2$*.

It is easy to see that neither of the two sets of potential states of the world is identical with those of Table 3. So it is possible to empirically differentiate whether the agent acted in order to feel warmer or in order to save some money. Notably, in $M_2$* the two variables *H* and *W* are independent from each other (this was not the case in $M_1$*). In fact, if the agent only cares about the bill, the state of the heating system and the weather are uncorrelated: the heating would be off with good weather as well as with bad weather. This conclusion could also be drawn by the graphical analysis of Figure 3, in which *H* and *W* are separated by a so-called collider[4].

---

[4] It could also be the case that the agent wants to have a warm temperature in her room, and that at the same time she does not want to spend too much on her electrical bill. This could be represented by a model in which the state of the heating *H* 'listens' to both the room temperature *T* and the electrical bill *B*, as both effects would be intended.

## 6. Finalistic confounding and a counterfactual notion of 'finality'

The reason to perform statistical teleological analysis is that mere observation of some behaviour and of its effects does not give a transparent access to the agent's intentions–even assuming a valid causal relationship linking behaviour and effects. Imagine that somebody goes to the cinema and there he meets a person he has not seen for many years. Was the agent's aim that of meeting the person? Second example: the agent opens the window, determining an increase in oxygen in the room, but also a decrease of temperature. Was the agent aiming for both effects? These situations exemplify two main families of what we could call *finalistic confounding*, that is, a relationship between causes and effects that we do not want to interpret as teleological.

The first example takes the graphical shape of *a chain*, in which an action produces an immediate effect and this effect in turn produces another further effect. The agent goes to the cinema and unexpectedly meets someone. The fact of going to the cinema is a causal condition for meeting that person there, but how can we decide whether he just wanted to watch the movie? This concerns the *depth* of the intention. We need data about many theatre screenings to give an answer and 'cut' the chain at the correct link. The second example takes the graphical shape of *a fork*, in which the action is connected to two effects, but only one of these two is intentional. The agent opens the window to get some fresh air inside the room, but this also means that the room gets colder. How can we decide which of the two effects was the aim of the action? This concerns the *breath* of the intention. We need data about many rooms and windows, in different seasons, to identify the correct branch.

When observing correlations in our data, therefore, we must interpret them through the eyes of a model capable of explaining them, not only causally but also by differentiating effects which are intended and others which are non intended (effects which are finalistically 'spurious'). It is the model itself–the formalisation of an hypothesis that we apply to the world–which guides us and gives meaning to the data, exactly as it happens with causal analysis. Different models represent different hypotheses on the observed behaviour.

Let us resume the general procedure of teleological explanation presented in this paper. In order to perform teleological analysis we start with a causal model that we assume is valid (that is, correctly describing the actual world as far as we can say). Then we select one variable of it that we want to interpret as an intervention[5]. To build a final model, we remove all inbound arrows to the node representing the manipulated variable in the causal model. This means that the action "becomes uncaused" and the corresponding node in the causal DAG becomes parentless. We complete the final model by selecting some variables that we believe are the intended effects of the intervention, and we reverse the direction of the arrows connecting these effects and the intervention. The intended effects of the action have to be chosen among the causal effects of the manipulated variable in the grounding causal model. We then try to identify the final model as we would do with a causal model. If the model is identified by observational data, we conclude that the variables selected as intended effects were the ends of the action (to the best of our knowledge, and in reference to our modelling assumptions, as usual).

The reasoning behind final models can be enounced as such: "If we assume that the value of $A$ was not due to any variable that is its antecedent in a causal model, can we then predict its value starting from some set of variables $B$ that are its descendents in the same causal model?" Causality can be seen as a constraint on the possible worlds that can get actual and so leave traces in observational data. Finalistic interpretation (or 'finality', with a neologism) is a further, higher-order, analogous constraint. It is important to understand that saying that $do(A)$ listens to its effect $B$ does not imply any inversion of causation between $B$ and $A$: $A$ keeps causing $B$ and it's very important for the agent that it does that. Cooked noodles do not turn the stove on.

Both constraints–causal and final–require to go beyond observable correlations and to reason counterfactually. Finality simply requires to go a little further in the calculus of potential states. If we see things in this way, finalistic interpretation is by no means "less real", anti-scientific or more obscure than the causal one. Whenever trying to explain something as a cause, we look to prove counterfactual statements like "If *A* didn't happen, then *B* would not have happened either", in short, "*do(A)* implies *B*". This short expression is already intimately counterfactual. Now, whenever trying to explain something as an action, we look to prove counterfactual statements like "If the agent didn't intend to obtain *B*, she would not have done *do(A)*". This latter statement cannot be reformulated with the first-order verb 'happen', as in the statement "If *B* didn't happen, then *A* would not have happened either". Instead, we could write it in short form with the expression "*want(B)* implies *do(A)*". This short expression is intimately counterfactual in a complex way. It's therefore surely much harder to convert second-order counterfactuals in descriptive assertions including only observable variables. We will not delve further into this for the present paper.

---

[5] The variable intervened on must not necessarily be a basic action, that is, a bodily movement or other things that one can do without doing something else (Danto) [8]. However, models of basic actions could be easier to conceive.

### 7. "That is an action" is not a teleological explanation

The choice of one variable as intervention is arbitrary. The analyst has to choose, *a priori*, not only one causal model through which to interpret the data, but one portion of the empirical world that is assumed as the result of an intervention. Let us imagine that I am watching a person turning the stove on. It is my choice to interpret that specific human behaviour as action, and not simply as a fact or event, while the effect of heat on water remains for me purely causal. In current philosophical terms, teleological analysis as it is presented here relies on an *intentional stance* as defined by Daniel Dennett [9]: action is a matter of perception and modelling. Such approach refrains from stating whether or not some behaviour 'really' is or is not intentional and free[6]. There is no 'real' finality in the world as there is no 'real' causality, all that matters are data and models. To perform teleological explanations one has just to accept that the measured data come from a "surgically modified" causal model, and so that when we *see B* we are actually seeing the effects of *do(A)* and not of *A*.

---

[6] An ontological view of goal-oriented behaviour is instead defended by Woodfield [10].

This means that teleological analysis can potentially answer the question "Which are the ends of this action?" but not the apparently more basic question "Is this an action?" In fact, to ontologically differentiate causal and teleological aspects of the world would demand a criterion for choosing whether a causal model or a final model is ultimately correct for explaining some phenomena. This ontological view of goal-oriented behaviour would imply that in the world some things are causally determined, while some others are not. And this would clearly go against what Reichenbach [4] wrote, making causal models unidentifiable, and as a consequence making final models unidentifiable too.

Following the spirit of Anscombe's thought [2], causal models and final models are different descriptions of the same variables. In any given context, the analyst can choose to see human behaviour as caused or as finalised. In the first case, the analyst will look for some causal variables explaining the observed behaviour in a causal model. For example, one can ask why people smoke by looking at remote causes in their biography, such as parents or friends who smoke. In the second case, the analyst cuts off all causal determinants of action and explains it only by reference to its effects in a final model. For example, one can ask why people smoke by looking at proximate effects, such as an increased level of pleasure signals in the brain.

This means that teleological explanation cannot be reduced to causal explanation. Causality and goal-oriented behaviour ('finality') are not two alternative ways in which the world works. They are different ways of describing and interpreting the world, and especially its measurable correlations. In this paper we assumed that causal interpretation is more basic than finalistic interpretation. No action can be defined and explained in a causal model, but any final model is based on an underlying causal model and its assumptions.

Instead of reducing intentions to (mental) causes, we suggest formalising causal and teleological relationships with the same concept: that of influence or, as Pearl writes more effectively, 'listening'. An effect *B* 'listens' to its cause *A*, and we represent it graphically as an arrow connecting the node *A* to the node *B* in a causal model. An action *do(A)* also 'listens' to its end *B*, and we represent it graphically as an arrow connecting the node *B* to the node *do(A)* in a corresponding final model. So, conceptually as well as methodologically, there is no reduction between causes and ends, but just a common conceptualisation (the 'listening' relationship) and interventions on variables working as a bridge between two kinds of models.

The analysis of goal-oriented behaviour proposed here is practically a way to separate dependency relationships on two planes, a causal one and a teleological ('final') one. Keeping dependency relationships on one same plane would instead create loops, and so prevent empirical research. If cigarettes increase serotonin, and people smoke in order to increase serotonin, one cannot represent the two facts on one causal DAG, because this would produce a cycle. But one can differentiate two DAGs, one causal and one final, making identification possible.

### 7. Can we do without final models?

In the preceding pages we aimed at showing why teleological statistical analysis is possible. But is it also necessary? Anscombe's perspective was famously criticised by Davidson [3]. For Davidson, intentions are mental causes, and this has one main consequence: causal explanations are sufficient to account for actions. Therefore, it would be interesting to understand whether final models can be expressed in an equivalent way with causal ones simply by including the agent's intention as a node.

Let us try to produce a causal model equivalent to the final model presented in Figure 2, in which the intentional effect of turning the heating on was to raise a room's temperature. Figure 4 and Table 5 show one possible way of expressing the causal reduction of the final model in Figure 2. Figure 4 adds an unobservable variable *I* standing for the intention of affecting the room's temperature, and differentiates two successive measurements of the room temperature, $T_0$ (before action) and $T_1$ (after action).

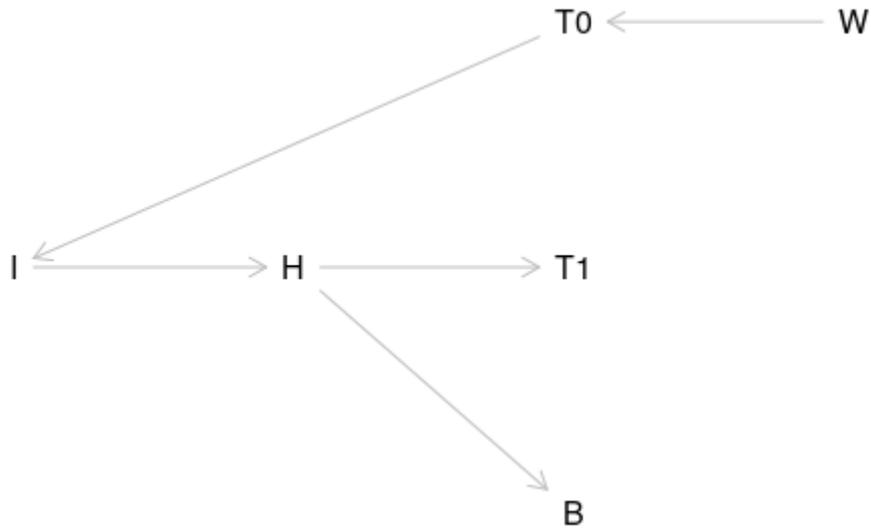

Figure 4: Graph representing a purely causal alternative to the model $M_1{}^*$

| W | $T_0$ | I | H | $T_1$ | B |
|---|---|---|---|---|---|
| 1 | 1 | 0 | 0 | 1 | 0 |
| 0 | 0 | 1 | 1 | 1 | 1 |

Table 5: The potential states of the world compatible with the model in Figure 4

The table reads as such: If the weather is good, then the temperature in the room is already warm, so the agent has no reason to turn the heating on; this results in the temperature of the room being warm and in the electrical bill staying cheap. If instead the weather is bad, then the temperature in the room is cold, so the agent is determined to turn the heating up; this results in the room becoming warm and the electrical bill to become expensive. The small world has only these two causally possible outcomes, which correspond to the ones we saw in Table 2.

From this perspective there is no difference anymore between *to rise* and *to raise*, that is, between a fact *H* and the result of an action *do(H)*. As a consequence, something else gets lost in the way: namely the very difference between causes and ends. In Figure 4, the intention *I* is a mediated effect of the external weather and a mediated cause of the amount on the electrical bill. So we could correctly say that "The heating is on because the weather is bad". However, is this what we honestly want to answer to the question "Why is the heating on?" or even more precisely to the question "Why has the heating been turned on?". Causal explanations cannot answer anything like "The heating is on because of its impact on the room temperature", and cannot account for the counterfactual "Had the agent liked the room temperature, she would have not turned on the heating". In other terms, there are some questions asked by the human and social sciences which could only receive a quantitative answer by models explicitly accounting for actions and their intentions in their very structure (and not simply by adding unobservable variables).

The counterfactual "Had the weather been good, then the heating would be off" is the only kind of answer compatible with the causal model in Figure 4. And it depicts heating as *H*, and not as *do(H)*. There can be no authentic *in order to* in a causal model, because causes do not have primary or preferential effects: all their effects are equal. The heating makes both the room warmer and the bill more expensive, without any 'qualitative' distinction.

To further prove the point, the intention *I* could be erased from the graph in Figure 4 without any loss of generality (the weather simply affects the room temperature, and so the state of the heating system), producing a smaller and equally valid causal model. It could even be possible to produce an even smaller and equally valid causal model by removing from the graph the measurement $T_0$, ending up with the model drawn in Figure 5.

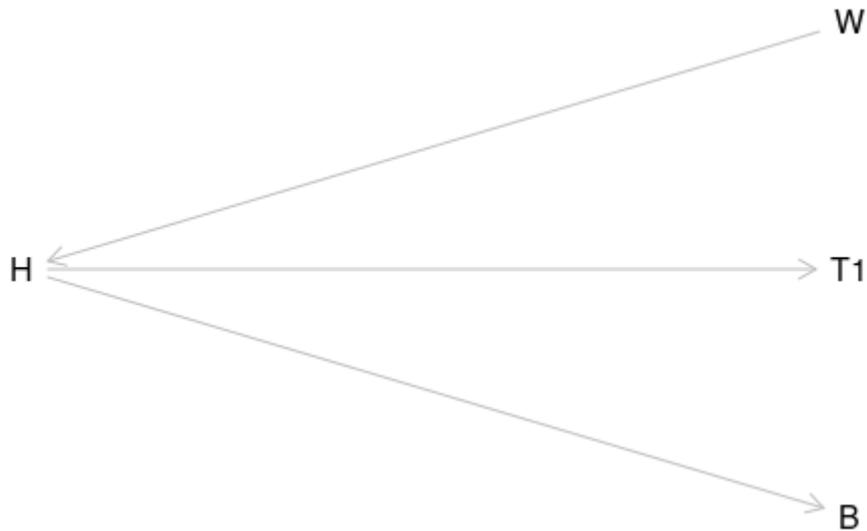

Figure 5: Graph representing another causal alternative to the model $M_1$*

In a purely causalist perspective, such as Davidson's, there are no interventions and no necessity for modelling intentions: they are just links of the global chain of causation, so they can be dissolved into other variables. And still, can we say that Figure 5 and Figure 2 show the same thing? From a statistical perspective, this is not the case: the nodes are connected differently in the two models. Here it is the weather that determines whether the heating is on or off (and not the intended temperature of the room, as in Figure 2). The electrical bill and the room temperature are two effects equivalent to the weather–but certainly not to the (absent) agent.

## 8. Conclusions

The paper has presented a way to address the finalistic interpretation of statistical correlations. It presented final models as an extension of (and not as an alternative to) causal models, including assumptions about interventions and their intended effects. It introduced one toy example of an identifiable final model, and briefly introduced the problem of finalistic confounding.

The next step is to generalise final models: Which kinds of models are identifiable? That is, which kinds of teleological *why* questions can be given an answer with these tools? The perspective presented in this paper is grounded on a counterfactual concept of goal-oriented behaviour or 'finality'. We aim to prove the validity of a teleological connection, that is, the truth of the counterfactual: "Had the agent not intended to produce *B*, she would not have done *A*". It could be interesting to delve further into this kind of counterfactual.

The author declares no conflict of interest. The author states no funding involved.

The diagrams have been obtained with the software *dagitty*: